\pdfoutput=1

\documentclass[11pt]{article}
\usepackage{enumitem}
\usepackage{amsmath}
\usepackage{amssymb}
\usepackage{algorithm}
\usepackage{algorithmic}
\usepackage{mathtools}

\usepackage[final]{acl}

\usepackage{times}
\usepackage{latexsym}

\usepackage[T1]{fontenc}

\usepackage[utf8]{inputenc}
\usepackage{multirow} 
\usepackage{microtype}

\usepackage{inconsolata}

\usepackage{graphicx}

\title{SelfRACG: Enabling LLMs to Self-Express and Retrieve for Code Generation}

\author{
  \textbf{Qian Dong}\textsuperscript{1,3},
  \textbf{Jia Chen}\textsuperscript{2},
  \textbf{Qingyao Ai}\textsuperscript{3,1}\thanks{\ \ Corresponding author.},
  \textbf{Hongning Wang}\textsuperscript{1},
  \textbf{Haitao Li}\textsuperscript{1},\\
  \textbf{Yi Wu}\textsuperscript{2},
  \textbf{Yao Hu}\textsuperscript{2},
  \textbf{Yiqun Liu}\textsuperscript{1},
  \textbf{Shaoping Ma}\textsuperscript{1}
\\
\textsuperscript{1}DCST, Tsinghua University, Beijing, China\\
\textsuperscript{2}Xiaohongshu Inc., China
\textsuperscript{3}Quan Cheng Laboratory, China
\\
\texttt{\{dq22,liht22\}@mails.tsinghua.edu.cn}, 
\texttt{\{aiqy,hw-ai,yiqunliu,msp\}@tsinghua.edu.cn},\\
\texttt{\{chenjia2,xiaohui,xiahou\}@xiaohongshu.com}
}

\begin{document}
\maketitle
\begin{abstract}
Existing retrieval-augmented code generation (RACG) methods typically use an external retrieval module to fetch semantically similar code snippets used for generating subsequent fragments. 
However, even for consecutive code fragments, the content often diverges due to logical progression, resulting in a content gap.  
This gap undermines the performance of current RACG methods, as \textit{external} retrieval modules based on content matching fail to infer the specific information need of LLMs to generate the next code fragment.
Therefore, we propose \textbf{SelfRACG}, a novel paradigm that enables large language models (LLMs) to \textbf{Self}-express their information needs to enhance \textbf{RACG}.
Specifically, SelfRACG includes an information need expression module and a two-stage information need-guided training strategy, which encourages LLMs to express their information need.
Extensive experiments demonstrate that SelfRACG can retrieve external knowledge that better aligns with the LLM's own information needs, resulting in superior generation performance compared to vanilla RACG.
Moreover, both the training and deployment costs for retrieval in our framework are much lower than those of the strongest retrieval model.
~\footnote{https://github.com/CSQianDong/SelfRACG}
\end{abstract}

\section{Introduction}

\begin{figure}[t]
    \centering
    \includegraphics[width=0.8\linewidth]{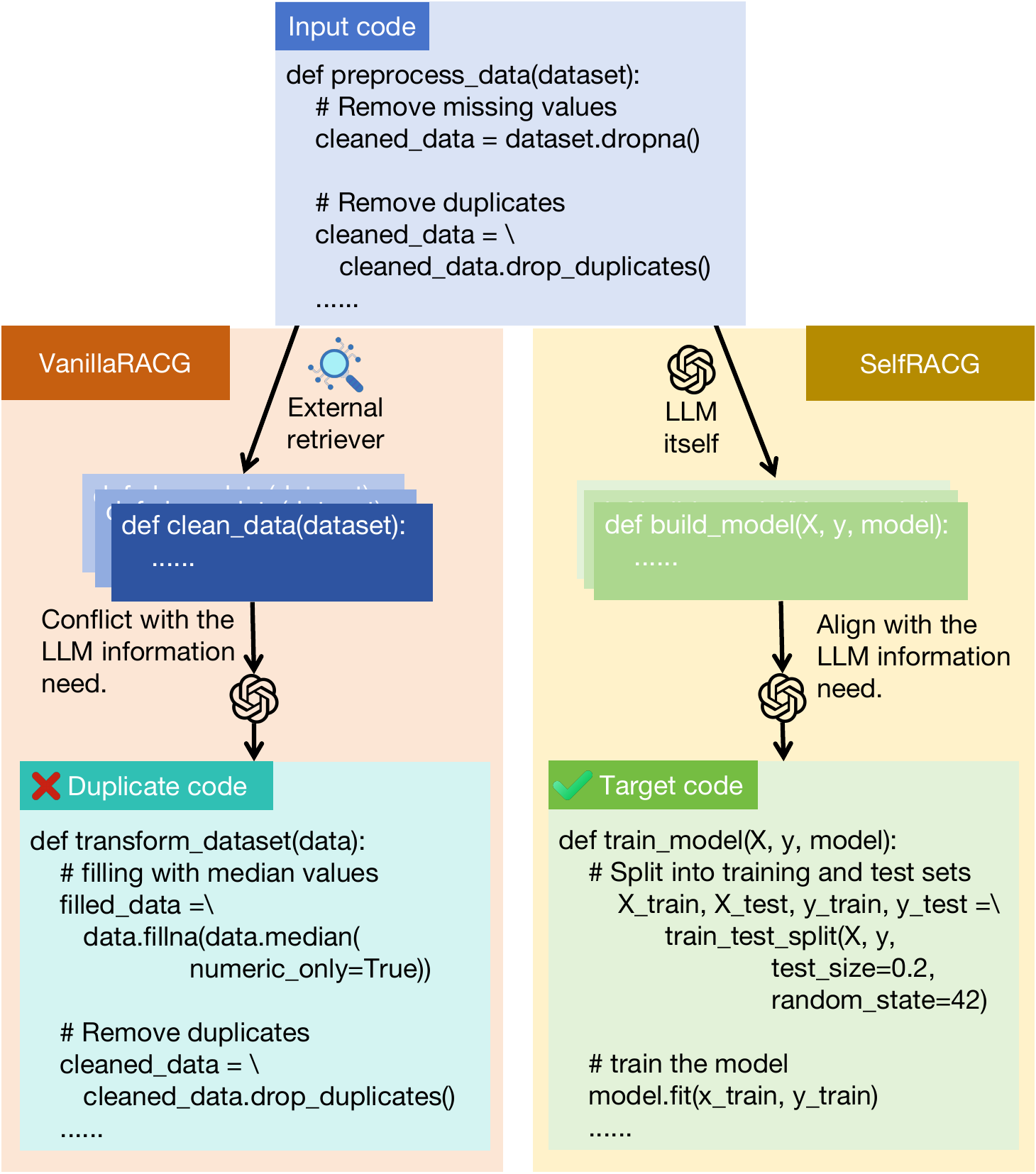}
    \caption{Comparison of vanilla RACG vs. SelfRACG.}
    \label{fig:comparison}
    \vspace{-10pt}
\end{figure}

Current retrieval-augmented code generation (RACG) techniques typically employ external retrieval modules, such as BM25~\cite{robertson2009probabilistic} or embedding-based retrieval models~\cite{muennighoff2024generative,lee2024nv}, to identify code fragments that closely align in content with preceding codes, thereby enhancing the generation of subsequent fragments.

However, even consecutive code fragments can differ significantly in content due to the progression of logic.
As shown in Figure~\ref{fig:comparison}, the preceding code fragment might implement a function for data pre-processing, such as cleaning or normalizing a dataset, while the subsequent fragment could define an entirely different function for training a machine learning model on the processed data. 
Although these two fragments are part of the same workflow, their content differs substantially, as each addresses a distinct functionality.
Consequently, existing RACG methods struggle to effectively improve overall performance, as the \textit{external} retrieval modules often fail to accurately fetch the knowledge required for code generation.
These external retrievers typically rely on similarity to retrieve code fragments for RACG. However, the retrieved fragments are often functionally similar to the existing code, rather than specifically aligned with the LLM's actual generation needs.
Building on the observation that LLMs inherently possess the capability to generate the next token, we argue that the hidden states associated with the next token already encapsulate the information needs required for future content generation. 
Therefore, a critical research question is: \textbf{Can LLMs express their own information needs for RACG?}


To enable LLMs to \textbf{Self}-express their information needs and perform \textbf{RACG}, our \textbf{SelfRACG} paradigm introduces two core components: an Information Need Expression (INE) module and a two-stage Information Need-Guided (ING) training strategy.
Specifically, INE module is achieved by a parameter-efficient fine-tuning technique named Layer-wise Low-Rank Adaptation (L-LoRA).
Through L-LoRA, we implement a retrieval-aware attention operation parallel to self-attention at each layer of the LLM. With minimal additional parameters and training costs, we extract the LLM's information needs from the hidden state of the next token.
Since the retrieval-aware attention operates completely in parallel with the LLM's original self-attention, the next-token hidden states remain unaffected.
Consequently, the original generation capabilities of the LLM can be fully preserved.
The second key technique of SelfRACG is ING, a two-stage training strategy that enables LLMs accurately retrieve the information required for subsequent code generation. 
The first stage of ING leverages existing code from GitHub to create training samples, providing the LLM with foundational code retrieval skills. The second stage then uses a small amount of LLM-synthesized data to further train the model, aligning its retrieval capabilities with its own generation preferences.

We conduct extensive experiments to validate our method on two comprehensive benchmarks using various top-tier code LLMs.
The experimental results demonstrate that SelfRACG effectively bridges the content gap between retrieved and subsequent code fragments, thereby improving the quality of code generation.
Moreover, both the training and deployment costs of our INE module for retrieval are much lower than those of the strongest retrieval module.

The key contributions of our work can be summarized as follows:

\begin{itemize}[leftmargin=*]
    \item We propose SelfRACG, a novel paradigm that enables LLMs to self-express their own information needs for RACG.
    \item We introduce the INE module and ING training strategy to equip LLMs with the capability of expressing information needs at a low cost, without compromising the LLMs' original capabilities.
    \item We conduct comprehensive experiments on two benchmarks using multiple code LLMs, demonstrating effectiveness of our method.
\end{itemize}

\section{Related Work}

\subsection{Code Large Language Models}
Recent advancements in code large language models (LLMs), such as Qwen2.5-Coder~\cite{hui2024qwen2}, OpenCoder~\cite{huang2024opencoder}, and DeepSeek-Coder~\cite{guo2024deepseek}, have demonstrated impressive capabilities in code generation. These models leverage vast amounts of code data to provide functional code snippets across diverse tasks. However, their performance often degrades when dealing with repository-level code completion, which requires fine-grained understanding of repository-specific contexts and dependencies~\cite{zhang2023repocoder,wang2024coderag}. 
To mitigate this issue, various retrieval-augmented generation methods have been proposed~\cite{zhang2023repocoder,liu2024graphcoder,lu2022reacc}.

\subsection{Retrieval-Augmented Code Generation}
RACG has emerged as a promising paradigm for enhancing code generation by integrating external context into the generation pipeline~\cite{tan2024prompt,parvez2021retrieval,gao2023retrieval}. 
Traditional RACG methods rely on retrievers like BM25~\cite{robertson2009probabilistic} or embedding-based models~\cite{muennighoff2024generative,lee2024nv} to fetch contextually relevant code snippets from repositories. For example, RepoCoder~\cite{zhang2023repocoder} employs an iterative retrieval-generation framework to iteratively refine retrieved code snippets. Similarly, Repohyper~\cite{phan2024repohyper} introduces a semantic graph-based retrieval method to capture broader contextual relationships within repositories.
RACG could address the limitations of parametric knowledge in LLMs.
Despite the advantage, existing RACG approaches often struggle to bridge the content gap between retrieved codes and the model's information needs, leading to suboptimal generation performance.

\subsection{LLM-based Embedding Models} 
Beyond generative tasks, LLMs have also been adapted for text embedding tasks, enabling them to serve as powerful embedding models. 
GritLM~\cite{muennighoff2024generative} unifies generative and embedding tasks within a single model through generative representational instruction tuning through extensive training costs. 
This approach enables GritLM to excel in both embedding and generation.
Besides, gte-Qwen~\cite{li2023towards}, instructor~\cite{su2022one} and NV-Embed~\cite{lee2024nv} extend embedding tasks with instruction tuning, enabling fine-grained control over embedding generation by explicitly encoding task-specific instructions. 
These works highlight the potential of LLMs in embedding tasks, presenting new possibilities for retrieval and semantic understanding.

\section{Method}
\label{sec:method}
\begin{figure*}[t]
    \centering
    \includegraphics[width=0.8\linewidth]{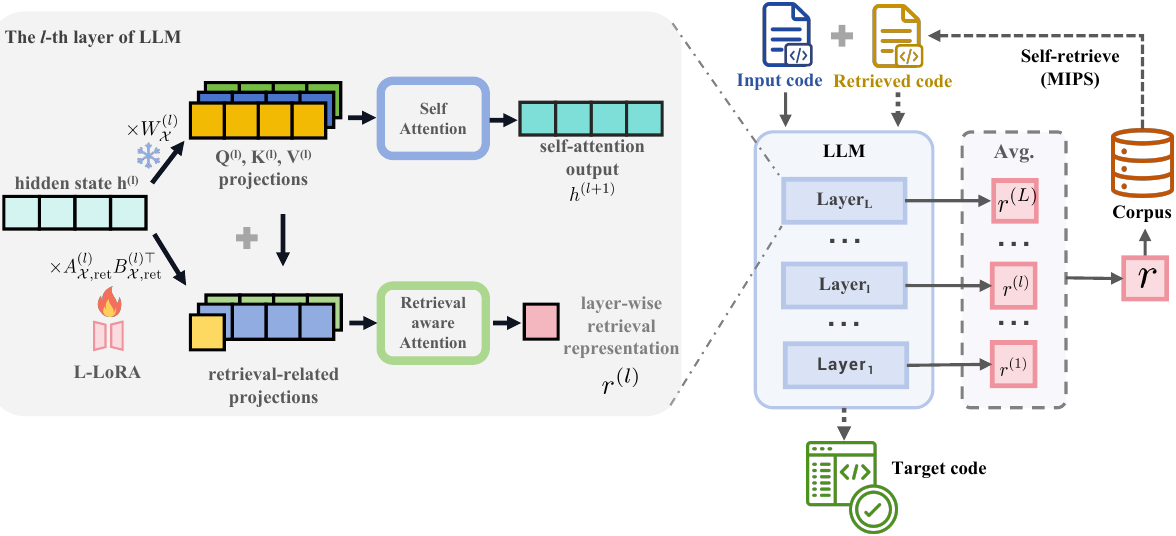}
    \caption{The workflow of SelfRACG. We use the $l$-th layer as an example and provide a detailed explanation of the relationship between retrieval-aware attention and self-attention within the gray box at the left of figure.}
    \label{fig:framework}
\end{figure*}

In this section, we introduce our proposed SelfRACG framework. 
Section~\ref{subsec:INE} introduces the Information Need Expression (INE) module, which enables LLMs to express their information needs from hidden states for retrieval.
Section~\ref{subsec:ING} outlines the two-stage Information Need-Guided (ING) training strategy, which is designed to train the INE module, ensuring that the retrieved code snippets align with the LLM's generation needs.
Figure~\ref{fig:framework} illustrates the workflow of SelfRACG.

\subsection{Information Need Expression Module}
\label{subsec:INE}
The design of the INE module is guided by two key principles: 
(1) introducing minimal additional parameters to reduce training costs, and (2) equipping the LLM to produce retrieval representations without compromising its original generation capabilities. To achieve this, the INE module is implemented as a parallel component to the traditional self-attention mechanism, incorporating a retrieval-aware attention operation through our proposed Layer-wise Low-Rank Adaptation (L-LoRA).

Specifically, for each transformer layer \( l \in [1, L] \), the projections $\mathcal{X}^{(l)}$ of original self-attention are first computed as
\begin{equation}
\label{eq:projections}
\mathcal{X}^{(l)} = h^{(l)}W_\mathcal{X}^{(l)},
\end{equation}
where $\mathcal{X}$ is a wildcard representing the $\{Q,K,V\}$ projections in self-attention.
$W_\mathcal{X}^{(l)} \in \mathbb{R}^{d \times d}$ are the weight matrices of the $\{Q,K,V\}$ projections.
Then, the scaled dot-product self-attention at layer $l$ is computed as
\begin{equation}
\label{eq:self-attn}
h^{(l+1)} = \text{Softmax}\left(\frac{Q^{(l)}{K^{(l)}}^\top}{\sqrt{d}}\right)V^{(l)}.
\end{equation}

To enable parameter-efficient tuning of LLMs, the vanilla LoRA~\cite{hu2021lora} decomposes weight updates $\Delta W_\mathcal{X}^{(l)}\rightarrow \alpha A_\mathcal{X}^{(l)}B_\mathcal{X}^{(l)\top}$. 
For each projection $\{Q,K,V\}$, the computation becomes

\begin{equation}
\label{eq:lora}
\Tilde{\mathcal{X}}^{(l)} = \mathcal{X}^{(l)} + \alpha h^{(l)}A_\mathcal{X}^{(l)}B_\mathcal{X}^{(l)\top},
\end{equation}
where $\alpha$ is a scaling factor that controls the magnitude of the adaptation, and $rank$ represents the rank of low-rank matrices $A_\mathcal{X}^{(l)}, B_\mathcal{X}^{(l)} \in \mathbb{R}^{d \times rank}, rank \ll d$. Subsequently, \( h^{(l+1)} \) is computed by Equation (\ref{eq:self-attn}) by replacing $\mathcal{X}$ with $\Tilde{\mathcal{X}}$.

Vanilla LoRA shifts the hidden states of the subsequent layers, potentially affecting the generation quality of the LLM. 
However, there is a significant gap between retrieval and generation tasks, making it challenging to optimize both simultaneously with limited training resources. 

Therefore, as illustrated in Figure~\ref{fig:framework}, our approach computes retrieval-related projections using L-LoRA and introduces a parallel attention mechanism alongside the original self-attention to produce representations specifically for the retrieval task.
The computation of retrieval-related projections is similar to vanilla LoRA, which can be defined as
\begin{equation}
\label{eq:ret_projections}
\mathcal{X}_{\text{ret}}^{(l)} = \mathcal{X}^{(l)} + \underbrace{\alpha h^{(l)} A_{\mathcal{X}, \text{ret}}^{(l)} B_{\mathcal{X}, \text{ret}}^{(l)\top}}_{\text{Extra computational cost.}}.
\end{equation}
Only \( \alpha h^{(l)} A_{X, \text{ret}}^{(l)} B_{X, \text{ret}}^{(l)\top} \) represents the additional computational cost of calculating the retrieval-related projections, while \( \mathcal{X}^{(l)} \) can be directly reused from the original self-attention, as shown in the left of Figure~\ref{fig:framework}.
Notably, we only require the query projection of the last token $t$ to compute the layer-wise retrieval representation.
The layer-wise retrieval representation can be computed by
\begin{equation}
\label{eq:retrieval-representation}
r^{(l)} = \text{Softmax}\left(\frac{{Q}_{ret,t}^{(l)}{{K}_{ret}^{(l)}}^\top}{\sqrt{d}}\right){V}_{ret}^{(l)},
\end{equation}
where \( {Q}_{ret,t}^{(l)} \) is query projection for last token \( t \). 

The information need of the LLM can be expressed as the mean pooling of each layer's retrieval representations, which can be computed as
\begin{equation}
\label{eq:mean-pooling}
r = \frac{1}{L} \sum_{l=1}^{L} r^{(l)}.
\end{equation}
This mean-pooling aggregation helps refine the retrieval process by leveraging both shallow and deep contextual features from self-attention.

L-LoRA and vanilla LoRA primarily differ in how the computed projections are used. After obtaining the retrieval-related projections $\mathcal{X}^{(l)}_{ret}$ through L-LoRA, they are employed in a parallel retrieval-aware attention operation alongside the original self-attention. In contrast, projections $\Tilde{\mathcal{X}}^{(l)}$ in Equation (\ref{eq:lora}) obtained through vanilla LoRA are used within the original self-attention operation. As a result, L-LoRA effectively decouples the generation and retrieval tasks, enabling the rich information encoded in the LLM's hidden states to be efficiently extracted for retrieval, while preserving the model's original generation capabilities.

\subsection{Information Need-Guided Training}  
\label{subsec:ING}  
\begin{figure}[t]
    \centering
    \includegraphics[width=0.8\linewidth]{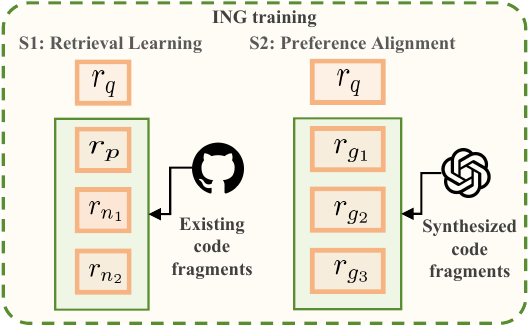}
    \caption{The illustration of two stage in ING training.}
    \label{fig:ING}
\end{figure}

To efficiently align the LLM's retrieval capability with its generation needs, we propose a two-stage training strategy, namely Information Need-Guided (ING) training, as illustrated in Figure~\ref{fig:ING}. 

\noindent\textbf{Stage 1: Retrieval Learning.}
In Stage 1 of ING training, we leverage existing GitHub repositories to construct extensive unsupervised training pairs. For a code fragment \( q \), the positive sample is the next code fragment \( p \), while the negative pool \( \mathcal{N} \) comprises code fragments from in-batch training pairs. 
After stage 1 training, the INE module gains the ability to capture general logical patterns, such as common design structures and typical coding habits.
However, it is not aligned with the LLM's generation preferences. 

\noindent\textbf{Stage 2: Preference Alignment.}
To align retrieval capability with the LLM's generation preference, we synthesize candidate fragments \( \{g_1, g_2, ..., g_k\} \) via LLM itself for a code fragment \( q \). The positive sample \( g_1 \) corresponds to the LLM's immediate next-step generated code fragment, while the negative pool \( \mathcal{N} \) combines synthetic negatives \( \{g_2, ..., g_k\} \) and in-batch negatives. This stage addresses permutation ambiguities by training the INE module to identify the precise fragment required for next-step generation.

Both stages employ the contrastive learning objective, which can be defined as
\begin{equation}  
\mathcal{L} = -\log \frac{e^{s(r_q, r_p)}}{e^{s(r_q, r_p)} + \sum_{r_n \in \mathcal{N}} e^{s(r_q, r_n)}},  
\end{equation}
where \( r_q \) denotes the context representation from INE. $s(\cdot,\cdot)$ means the inner production operation.
\( r_p \) and \( r_n \)  are the positive and negative candidate representations, respectively.
Following prior works~\cite{muennighoff2024generative,lee2024nv,li2023towards},  when calculating \( r_p \) and \( r_n \) through INE, we prepend a prefix to denote them as retrieval candidates.
In contrast, \( r_q \) is computed directly, without any prefix or suffix. This distinction ensures that the representation of the context and the candidates are produced in different formats.


In summary, the INE module and the ING training strategy enable LLMs to express their information needs and retrieve code fragments that are useful for subsequent generation at low computational cost.

\subsection{Inference}

To effectively leverage the retrieval representations produced by the INE module during inference, we follow a straightforward yet efficient process that ensures efficiently code retrieval for RACG.

First, we pre-process the code repository by encoding each code fragment using the INE module. 
These representations are then stored in a vector database, creating a dense embedding index that can be efficiently searched at inference time.

During inference, the incomplete code fragment is first processed by the same INE module to generate its retrieval representation. This representation is then used to perform Maximum Inner Product Search (MIPS)~\cite{karpukhin2020dense}, allowing LLM to identify the code snippets that align with its generation needs. 
The inference process is illustrated on the right side of Figure~\ref{fig:framework}. 
The solid lines represent the process by which the LLM expresses its information needs, while the dashed lines denote the subsequent step where the retrieved code fragments (via MIPS) are combined with the original code fragment to form a new input, which is then used to generate the target code.

\section{Experimental Setup}

\subsection{Baselines}
The improvement of the SelfRACG paradigm lies in integrating the retrieval module with the LLM and ensuring that the retrieved results align closely with the LLM's information needs. Therefore, we validate the effectiveness of our method by comparing it with different retrieval models. 
We refer to the RACG paradigm with an additional retrieval model as VanillaRACG. For VanillaRACG, we use the following retrieval models:

\begin{itemize}[leftmargin=*]
    \item \textbf{BM25}~\cite{robertson2009probabilistic}: A traditional sparse retrieval method based on term frequency-inverse document frequency. 
    \item \textbf{OpenAI-Small/Large}~\footnote{https://platform.openai.com/docs/api-reference/embeddings}: OpenAI-Small/Large are the two strongest closed-source embedding models from OpenAI, specifically \texttt{text-embedding-3-small} and \texttt{text-embedding-3-large}. These models can be used for effective code fragment retrieval. 
    \item \textbf{GritLM}~\cite{muennighoff2024generative}: 
    GritLM is the closest baseline to SelfRACG, as it enables LLMs to support both embedding and generation tasks simultaneously. However, it relies on full parameter fine-tuning, which requires substantial computational resources and training data.
    \item \textbf{NV-Embed-v2}~\cite{lee2024nv}: A state-of-the-art embedding model ranked top on the MTEB~\cite{muennighoff2022mteb} benchmark.
    However, NV-Embed-v2 can only be used to produce embeddings and is not capable of handling generation tasks.
\end{itemize}

\subsection{Benchmarks}
We evaluate our method on two benchmarks:

\begin{itemize}[leftmargin=*]
    \item \textbf{RepoEval}~\cite{zhang2023repocoder}: This benchmark is designed for repository-level code completion tasks and comprises three sub-tasks:
    \begin{itemize}[leftmargin=*]
        \item \textbf{Line-level Completion}: Tasks that require the model to complete a single line of code based on the context.
        \item \textbf{API-level Completion}: Tasks that involve completing function calls or API invocations.
        \item \textbf{Function-level Completion}: Tasks that need the model to complete the body of a function based on its definition or usage.
    \end{itemize}
    \item \textbf{CrossCodeEval}~\cite{ding2024crosscodeeval}: This benchmark focuses on cross-file contextual understanding. Built upon real-world code repositories, it challenges models to utilize dependencies and contextual information across multiple files to generate accurate code completions.
\end{itemize}

\subsection{Evaluation Metrics}
We adopt multiple metrics to comprehensively evaluate both retrieval and generation performances:

\subsubsection{Code Completion Metrics}
\begin{itemize}[leftmargin=*]
    \item \textbf{Exact Match (EM)}: Measures the percentage of predictions that exactly match the ground-truth code. This metric is applied to all sub-tasks.
    \item \textbf{Edit Similarity (ES)}: Evaluates the similarity between the generated and ground-truth code fragments using a normalized Levenshtein distance. It is also employed for all sub-tasks.
    \item \textbf{Pass@1}: It assesses whether the code generated via greedy decoding passes all relevant test cases when replacing the original code. This metric is exclusively utilized for function-level completion task in RepoEval~\cite{zhang2023repocoder}.
\end{itemize}

\subsubsection{Code Retrieval Metrics}
\begin{itemize}[leftmargin=*]
    \item \textbf{Recall@K}: Evaluates the proportion of ground-truth code fragments that are successfully retrieved within the top K results.
    \item \textbf{MRR@10}: Mean Reciprocal Rank at 10, calculated as the mean of the reciprocal ranks of the ground-truth code fragment for each sample. The reciprocal rank is given by \( \frac{1}{pos} \), with \( pos \) being the position of the ground-truth code fragment. 
\end{itemize}

\begin{table*}[t] 
\caption{The code completion experimental results of OpenCoder-Base} 
\resizebox{\textwidth}{!}{
\begin{small}
\begin{tabular}{l|cc|cc|ccc|cc|cc} 
\hline
& \multicolumn{2}{c|}{RepoEval-API} & \multicolumn{2}{c|}{RepoEval-Line} & \multicolumn{3}{c|}{RepoEval-Function} & \multicolumn{2}{c|}{CrossCodeEval} & \multicolumn{2}{c}{MEAN} \\ 
\cline{2-12} 
& EM & ES & EM & ES & EM & ES & Pass@1 & EM & ES & EM & ES \\ 
\hline 
\multicolumn{12}{c}{OpenCoder-1.5B-Base} \\ 
\hline 
w/o Retrieval & 0.258 & 0.557 & 0.344 & 0.603 & 0.050 & 0.388 & 19.8\% & 0.033 & 0.421 & 0.171 & 0.492 \\ 
\hline 
VanillaRACG &  &  &  &  &  &  & &  & &  &  \\ 
\multicolumn{1}{r|}{\footnotesize\textit{w/ BM25}} & 0.297 & 0.604 & 0.396 & 0.636 & 0.065 & 0.414 & 24.8\% & 0.055 & 0.449 & 0.203 & 0.526 \\ 
\multicolumn{1}{r|}{\footnotesize\textit{w/ OpenAI-small}} & 0.303 & 0.607 & 0.408 & 0.647 & 0.074 & \underline{0.430} & 24.8\% & 0.074 & 0.457 & 0.215 & 0.535 \\ 
\multicolumn{1}{r|}{\footnotesize\textit{w/ OpenAI-large}} & 0.303 & 0.611 & 0.415 & 0.652 & \underline{0.084} & 0.421 & 26.3\% & 0.072 & 0.454 & 0.218 & 0.534 \\ 
\multicolumn{1}{r|}{\footnotesize\textit{w/ NV-Embed-v2}} & \underline{0.328} & \underline{0.626} & {0.423} & 0.650 & 0.074 & 0.415 & 26.9\% & 0.077 & 0.463 & 0.225 & 0.538 \\ 
\multicolumn{1}{r|}{\footnotesize\textit{w/ GritLM-7B}} & 0.322 & 0.624 & \underline{0.428} & \underline{0.657} & {0.081} & {0.427} & \underline{27.6\%} & \underline{0.086} & \textbf{0.470} & \underline{0.229} & \underline{0.545} \\ 
\hline
SelfRACG & \textbf{0.342} & \textbf{0.636} & \textbf{0.439} & \textbf{0.668} & \textbf{0.093} & \textbf{0.439} & \textbf{29.4\%} & \textbf{0.088} & \underline{0.465} & \textbf{0.241} & \textbf{0.552} \\ 
\hline 
\multicolumn{12}{c}{OpenCoder-8B-Base} \\ 
\hline 
w/o Retrieval & 0.292 & 0.591 & 0.377 & 0.628 & 0.059 & 0.414 & 24.8\% & 0.048 & 0.437 & 0.194 & 0.518 \\ 
\hline
VanillaRACG &  &  &  &  &  &  & &  & &  &  \\ 
\multicolumn{1}{r|}{\footnotesize\textit{w/ BM25}} & 0.346 & 0.633 & 0.444 & 0.675 & 0.081 & 0.445 & 29.1\% & 0.084 & 0.477 & 0.238 & 0.557 \\ 
\multicolumn{1}{r|}{\footnotesize\textit{w/ OpenAI-small}} & 0.349 & 0.633 & 0.443 & 0.675 & 0.090 & \underline{0.467} & 29.4\% & 0.104 & 0.483 & 0.247 & 0.565 \\ 
\multicolumn{1}{r|}{\footnotesize\textit{w/ OpenAI-large}} & 0.338 & 0.631 & 0.450 & {0.685} & \underline{0.099} & {0.464} & \underline{31.9}\% & 0.106 & 0.487 & 0.248 & 0.567 \\ 
\multicolumn{1}{r|}{\footnotesize\textit{w/ NV-Embed-v2}} & \underline{0.365} & {0.648} & \underline{0.469} & 0.685 & 0.087 & 0.459 & 29.1\% & 0.114 & {0.494} & {0.259} & {0.572} \\ 
\multicolumn{1}{r|}{\footnotesize\textit{w/ GritLM-7B}} & 0.360 & \underline{0.650} & 0.468 & \underline{0.689} & \underline{0.099} & {0.463} & {30.7\%} & \underline{0.128} & \underline{0.501} & \underline{0.264} & \underline{0.576} \\ 
\hline
SelfRACG & \textbf{0.391} & \textbf{0.680} & \textbf{0.481} & \textbf{0.705} & \textbf{0.124} & \textbf{0.481} & \textbf{33.8\%} & \textbf{0.127} & \textbf{0.507} & \textbf{0.281} & \textbf{0.593} \\ 
\hline
\end{tabular}
\end{small}
}
\label{tab:opencodergen}
\vspace{-3mm}
\end{table*}

\subsection{Specific Settings}

To thoroughly evaluate the effectiveness of our method, we select multiple top-tier code LLMs across various sizes and versions.
We use the instructed version of Qwen2.5-Coder and~\cite{hui2024qwen2} the base version of OpenCoder~\cite{huang2024opencoder}. 
For Qwen2.5-Coder, we experiment with both 3B and 7B parameter sizes, and for OpenCoder, we use the 1.5B and 8B parameter sizes.

For the Stage 1 of ING, the code data is taken from the GitHub provided by CodeRAG-Bench~\cite{wang2024coderag}, which contains about 1.7 million code files. We filtered out repositories that are already used in benchmarks as well as files containing less than 20 lines of code. After filtering, we constructed a final corpus of about 1.3 million files. Following prior work~\cite{zhang2023repocoder,liu2024graphcoder}, we split the corpus into code fragments with a 20-line interval. The same splitting method was applied to the code in benchmarks. 
Notably, we did not use any supervised training datasets~\cite{xie2023t2ranking,nguyen2016ms,chen2025qilin}, which are widely adopted for training embedding models in prior works~\cite{dong2023i3,li2023sailer}. 

In the L-LoRA module, the rank is set to 16, and the alpha parameter is set to 32. For synthetic data generation, we use the vLLM~\cite{kwon2023efficient} inference library for acceleration. Starting with the first 20 lines of the preprocessed code corpus, we prompt the LLM to generate and complete the subsequent code fragments. In total, we generated 500k synthetic samples for the Stage 2 of ING.

We run all experiments on 8 NVIDIA A100 80GB GPUs. 
By utilizing multi-GPU parallelism, the largest model for synthetic data generation takes approximately 3.5 hours. The total data generation and training time for the largest model in the second phase of ING is 48 hours, corresponding to 384 GPU hours. 
In the Stage 2 of ING, we use 4 synthetic negative samples per code fragment.

\begin{table*}[htbp]
\caption{Retrieval performance of different retrieval models.}
\resizebox{\textwidth}{!}{
\begin{footnotesize}
\begin{tabular}{l|cc|cccc|c}
\hline
Model & \#Params & \#Dims & Recall@1 & Recall@3 & Recall@5 & Recall@10 & MRR@10 \\
\hline
External Retriever & & & &  &  &  &  \\
\multicolumn{1}{r|}{\footnotesize\textit{OpenAI-small}} & - & 1536 & 0.136 & 0.359 & 0.473 & 0.626 & 0.291 \\
\multicolumn{1}{r|}{\footnotesize\textit{OpenAI-large}} & - & 3072 & 0.149 & 0.359 & 0.484 & 0.644 & 0.304 \\
\multicolumn{1}{r|}{\footnotesize\textit{GritLM-7B}}  & 7B & 4096 & 0.174 & 0.444 & 0.592 & 0.748 & 0.356 \\
\multicolumn{1}{r|}{\footnotesize\textit{NV-Embed-v2}} & 7B & 4096 & 0.178 & 0.465 & 0.584 & 0.712 & 0.357 \\
\hline
SelfRACG & & & &  &  &  &  \\
\multicolumn{1}{r|}{\footnotesize\textit{OpenCoder-1.5B}} & 1.5B & 1536 & 0.208 & 0.433 & 0.528 & 0.664 & 0.358 \\
\multicolumn{1}{r|}{\footnotesize\textit{Qwen2.5-Coder-3B}} & 3B & 1536 & 0.239 & 0.517 & 0.621 & 0.748 & 0.411 \\
\multicolumn{1}{r|}{\footnotesize\textit{Qwen2.5-Coder-7B}} & 7B & 3584 & \textbf{0.244} & 0.508 & 0.616 & 0.746 & \textbf{0.413} \\
\multicolumn{1}{r|}{\footnotesize\textit{OpenCoder-8B}} & 8B & 4096 & 0.237 & \textbf{0.524} & \textbf{0.642} & \textbf{0.756} & \textbf{0.413} \\
\hline
\end{tabular}
\end{footnotesize}
}
\label{tab:retrieval}
\end{table*}
\section{Experimental Results}
\subsection{Code Completion Performance}
\label{subsec:codegen-performance}
In this section, we compare the code completion performance of VanillaRACG at different retrieval settings with SelfRACG.
The results are presented in Table~\ref{tab:opencodergen} and Table~\ref{tab:qwen2gen}.
As the trends in Table~\ref{tab:opencodergen} and Table~\ref{tab:qwen2gen} are similar, we have moved Table~\ref{tab:qwen2gen} to Appendix~\ref{sec:qwen25}.
From these tables, we can draw the following conclusions:
\begin{itemize}[leftmargin=*]
    \item Compared to no retrieval, RACG significantly improves the generation performance. Even the simplest retrieval model, BM25, leads to substantial improvements across all code completion sub-tasks and metrics.
    \item Embedding-based retrieval models outperform BM25, further enhancing code completion performance. This indicates that embedding-based retrieval models have stronger expressive capabilities than statistical methods, making them better at capturing the information needs of LLMs.
    \item LLM-based retrieval models deliver better generation results, highlighting the superior understanding power of LLMs in capturing complex information needs.
    \item Among all VanillaRACG configurations, GritLM-7B achieves the best generation performance. Fine-tuned on a large dataset, it is designed to integrate both retrieval and generation capabilities into one LLM. Compared to NV-Embed-v2, which focuses solely on the retrieval task, the data retrieved by GritLM-7B align more closely with the information needs of generation, resulting in better performance. 
    However, GritLM-7B requires significant computational resources to integrate both generation and embedding capabilities, according to its paper~\cite{muennighoff2024generative}. In contrast, SelfRACG outperforms GritLM-7B while using 1/8 of the GPU hours.
    \item SelfRACG outperforms VanillaRACG based on GritLM-7B. 
    This performance gain can be attributed to its unified design. SelfRACG uses the same LLM for both retrieval and generation, allowing it to extract retrieval representations directly from the next-token hidden states. This approach results in a closer alignment between the retrieved information and the LLM's own generation needs.
\end{itemize}

\subsection{Code Retrieval Performance}
In this subsection, we evaluate the retrieval performance of various retrieval models. 
Since both API-level and Function-level completions involve the permutation of implementations, it is difficult to evaluate the retrieval results. 
Theoretically, given the context, the line-level ground truth and the LLM’s information need for generating the next step are consistent. 
Therefore, we use the line-level completion samples from RepoEval for retrieval performance evaluation. 
Specifically, we treat the context as the query and the code fragment containing the target line as the positive ground truth, using Recall@K and MRR@10 to assess the retrieval performance.
The results are summarized in Table~\ref{tab:retrieval}. 
From the table, we observe the following trends: 
(1) SelfRACG outperforms other retrieval models in terms of MRR@10 and Recall@1. 
This indicates that by aligning the retrieval with the generation task, SelfRACG's retrieval capabilities are consistent with the generation logic of LLM, leading to more refined ranking results. 
(2) Among the different settings of SelfRACG, we observe that larger model sizes generally yield better retrieval performance.
Scaling up the model in terms of parameters may further enhance its ability to retrieve information that aligns with the generation preferences of the LLM.

\subsection{Alternative Retrieval Strategies}
\begin{table}[]
    \caption{Retrieval strategies comparison.}
    \resizebox{\linewidth}{!}{
    \begin{small}
    \begin{tabular}{l|c|c|c}
    \hline
    \textbf{Retrieval Strategy} & \textbf{API} & \textbf{Line} & \textbf{Func} \\
    \hline
    VanillaRACG (GritLM) & 0.650 & 0.690 & 0.463 \\
    \multicolumn{1}{r|}{\footnotesize\textit{w/ next fragment}} & 0.657 & 0.692 & 0.472 \\
    \multicolumn{1}{r|}{\footnotesize\textit{w/ next query}} & 0.673 & 0.705 & 0.476 \\
    SelfRACG & 0.680 & 0.705 & 0.481 \\
    \hline
    \end{tabular}
    \end{small}
    }
    \label{tab:retrieval_strategy_comparison}
    \vspace{-5mm}
\end{table}
We evaluate the generation performance of different retrieval strategies using the three levels (API, line and function) of completion tasks from RepoEval and the ES metric.
The base model used in this experiment is OpenCoder-8B-Base.
The comparison results are presented in Table~\ref{tab:retrieval_strategy_comparison}.
Directly using similar code fragments to augment in generating the next code segment yields the weakest performance.
The \textit{w/ next fragment} strategy directly uses the subsequent code fragment following the retrieved code fragment as external knowledge for RACG. 
However, the improvements are limited, as the subsequent fragment may not always align with the LLM's generation needs. 

The \textit{w/ next query} strategy leverages the LLM itself to generate a explicit query, which is then used to retrieve external knowledge by GritLM. As shown in Table~\ref{tab:retrieval_strategy_comparison}, this strategy achieves better results, as the LLM-generated queries are more aligned with its own generation needs. 
However, it still performs worse than SelfRACG on API and function-level completion tasks.
This is because SelfRACG directly produces a implicit query embedding from the next-token hidden states, which is more expressive than an explicit query.
Additionally, it avoids the extra step of converting an explicit query into an embedding via GritLM, reducing information loss. 
Therefore, SelfRACG achieves the best generation performance.

It is worth noting that the performance of different strategies on line-level tasks is relatively similar. This is because line-level completions are generally more deterministic given the context, making the LLM's information needs easier to capture compared to more complex API or function-level tasks.

\subsection{Ablation Studies}
\begin{table}[]
    \caption{Ablation studies of SelfRACG.}
    \resizebox{\linewidth}{!}{
    \begin{small}
    \begin{tabular}{l|cc|cc}
    \hline
    \multirow{2}{*}{Method} & \multicolumn{2}{c|}{OpenCoder-1.5B} & \multicolumn{2}{c}{OpenCoder-8B} \\
    \cline{2-5}
     & \textbf{MRR@10} & \textbf{ES} & \textbf{MRR@10} & \textbf{ES} \\
    \hline
    SelfRACG          & 0.358  & 0.6678 & 0.413  & 0.7048 \\
    \multicolumn{1}{r|}{\footnotesize\textit{w/o Stage 2.}}     & 0.316  & 0.6531 & 0.364  & 0.6912 \\
    \multicolumn{1}{r|}{\footnotesize\textit{w/o L-LoRA}}       & 0.296  & 0.6518 & 0.338  & 0.6847 \\
    \hline
    \end{tabular}
    \end{small}
    }
    \label{tab:ablation}
    \vspace{-5mm}
\end{table}

We conduct a step-by-step ablation of SelfRACG to analyze the impact of key techniques on retrieval and generation performance. Due to limited computational resources, we performed ablation experiments on OpenCoder using the RepoEval line-level completion data. The results are presented in Table~\ref{tab:ablation}. The first row shows the performance of SelfRACG using all techniques.

From the table, we can draw the following conclusions: The preference alignment in stage 2 of ING plays a crucial role in the model's performance. Without alignment with the generation preferences, knowledge conflicts occur, which negatively affect the model’s generation quality.
In the ablation experiment for L-LoRA, we fine-tuned the corresponding OpenCoder for embedding tasks using full parameter tuning. However, due to computational limitations, full parameter fine-tuning results in smaller batch sizes. In the full parameter fine-tuning setup, the batch sizes for 1.5B and 8B OpenCoder are 32 and 4, respectively. 
A smaller batch size leads to fewer negative samples in contrastive learning, thus decreasing the retrieval performance. Therefore, under limited resources, L-LoRA could achieve better performance.

\subsection{Overhead Comparison}
In this section, we present a comparison between SelfRACG and Vanilla RACG with an LLM-based embedding model, analyzing their overhead from both training and deployment perspectives.

\begin{table}[]
    \caption{Overhead comparison of different methods.}
    \resizebox{\linewidth}{!}{
    \begin{small}
    \begin{tabular}{l|c|c}
    \hline
    \textbf{Method} & \textbf{Training} & \textbf{Deployment} \\
    \hline
    GritLM-7B &  3,072 GPU hours & 14GB \\
    NV-Embed-v2 & - & 14GB \\
    SelfRACG & 384 GPU hours & 0.02GB \\
    \hline
    \end{tabular}
    \end{small}
    }
    \label{tab:overhead_comparison}
    \vspace{-5mm}
\end{table}

Table~\ref{tab:overhead_comparison} presents a comparison of training and deployment overhead for different methods. For a fair comparison, the results for SelfRACG are reported using Qwen-2.5-Coder-7B. 
Notably, GritLM-7B requires substantial training overhead. It is designed to handle both generation and embedding tasks simultaneously, requiring 8 nodes with 8 NVIDIA A100 80GB GPUs.
NV-Embed-v2 did not report its exact training resource consumption. 
SelfRACG based on Qwen-2.5-Coder-7B requires only a single node with 8 NVIDIA A100 80GB GPUs and a total of 384 GPU hours, equipping the LLM with self-expressive information need capabilities without collapsing its generation ability.
Both NV-Embed-v2 and GritLM-7B are standalone embedding models, requiring an additional 14GB of VRAM for deployment in half-precision mode. 
In comparison, SelfRACG shares the same LLM backbone as the generation model, with only a small number of additional parameters introduced.

\section{Conclusion}

In this work, we introduced SelfRACG that allows LLMs to self-express their own information needs for code generation. 
SelfRACG integrates an information need expression module and a two-stage information need-guided training strategy, improving retrieval performance while preserving generation capabilities.
Experiments on two benchmarks show that SelfRACG outperforms existing retrieval models at less than 1/8 of the training cost, providing a more efficient solution for RACG.

\section*{Acknowledgments}
This work is supported by the Key R\&D Program of Shandong Province (SYS202201) and Xiaohongshu Inc. 

\bibliography{custom}
\appendix
\section{Limitation}
We acknowledge several limitations in this study and aim to address them in future work. First, due to limited computational resources, our method has been validated only on LLMs with up to 8B parameters. Second, our approach has been evaluated exclusively on the RACG task; in future work, we plan to extend it to general retrieval-augmented generation scenarios to validate its broader applicability. In fact, the potential of L-LoRA extends beyond retrieval tasks. We will explore its application to various embedding tasks, enabling low-cost embedding capabilities for every LLM.

\section{Legal and Ethical Considerations}
In our research, ethical considerations are given top priority throughout the entire development process. We ensure that our work is free from any discriminatory elements and does not violate personal privacy. We are committed to using only publicly accessible and authorized datasets, avoiding any data that could lead to biased or harmful outcomes. All models, datasets, and code associated with this research are publicly available, enhancing transparency and enabling the community to conduct further research and validation.

\begin{figure*}[htbp]
    \includegraphics[width=\linewidth]{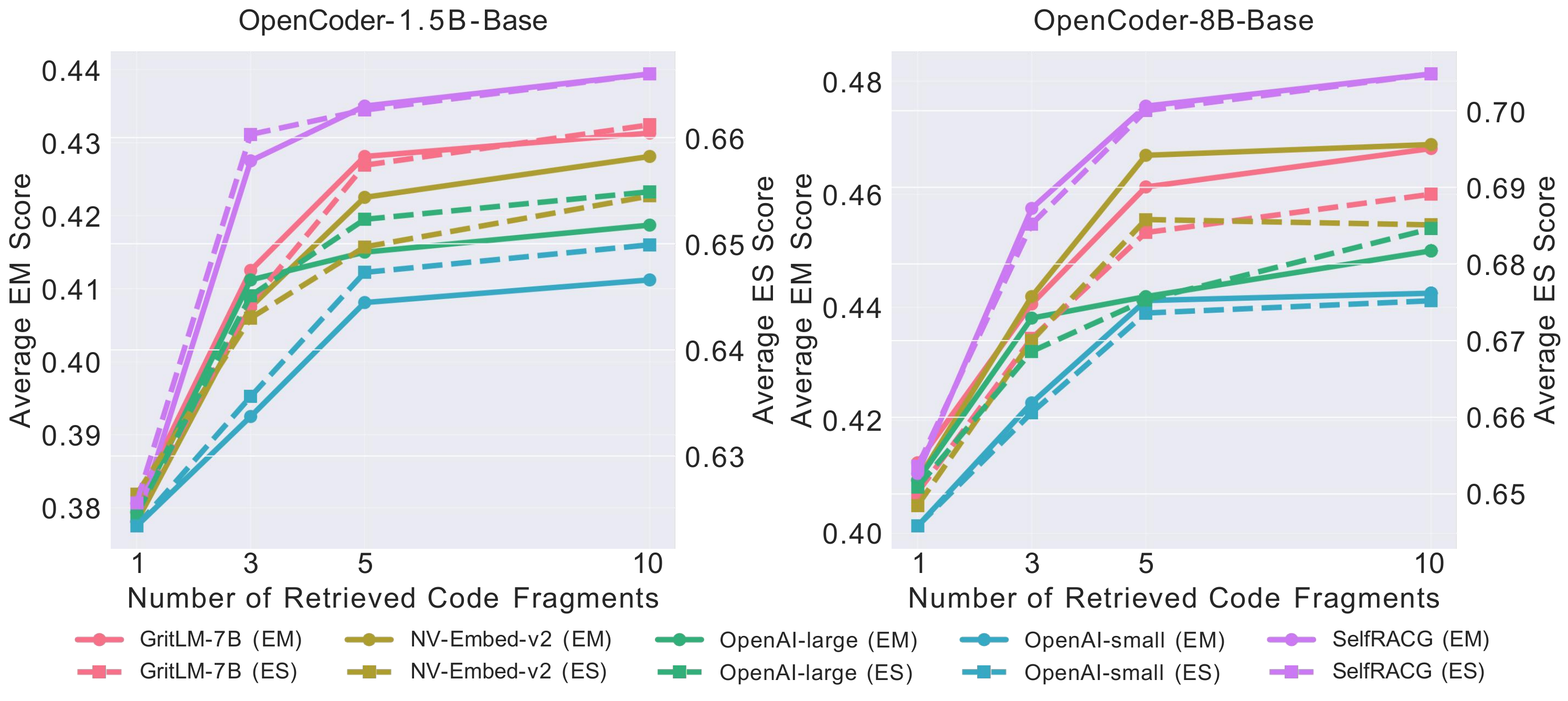}
    \caption{The generation performance with different number of retrieved code fragments.}
    \label{fig:docnums}
\end{figure*}

\section{Generation Performance with Different Number of Fragments}
In this section, we analyze the impact of the number of retrieved code fragments on the generation performance for the line-level completion task, and present the results in Figure~\ref{fig:docnums}. By jointly analyzing Table~\ref{tab:retrieval} and Figure~\ref{fig:docnums}, we can draw the following conclusions.
Although OpenCoder-1.5B has a lower Recall@10 compared to GritLM-7B and NV-Embed-v2, it performs better in terms of Recall@1 and MRR@10. This indicates that SelfRACG is able to rank relevant code fragment higher, leading to better generation results.
Due to the "lost-in-the-middle" issue~\cite{liu2024lost}, LLMs tend to ignore information from the middle of the context. 
Thus, a more accurate relevance modeling can significantly improve generation performance, as useful information is placed at the beginning of the context.
As shown in the left part of Figure~\ref{fig:docnums}, SelfRACG benefits from an increasing number of recalled code fragments, with performance improving rapidly as more fragments are included. When only five code fragments are used, SelfRACG using OpenCoder-1.5B and OpenCoder-8B already outperforms other retrieval models that use ten fragments.

\section{Prompt Template}
\begin{figure*}[t]
    \centering
    \includegraphics[width=\linewidth]{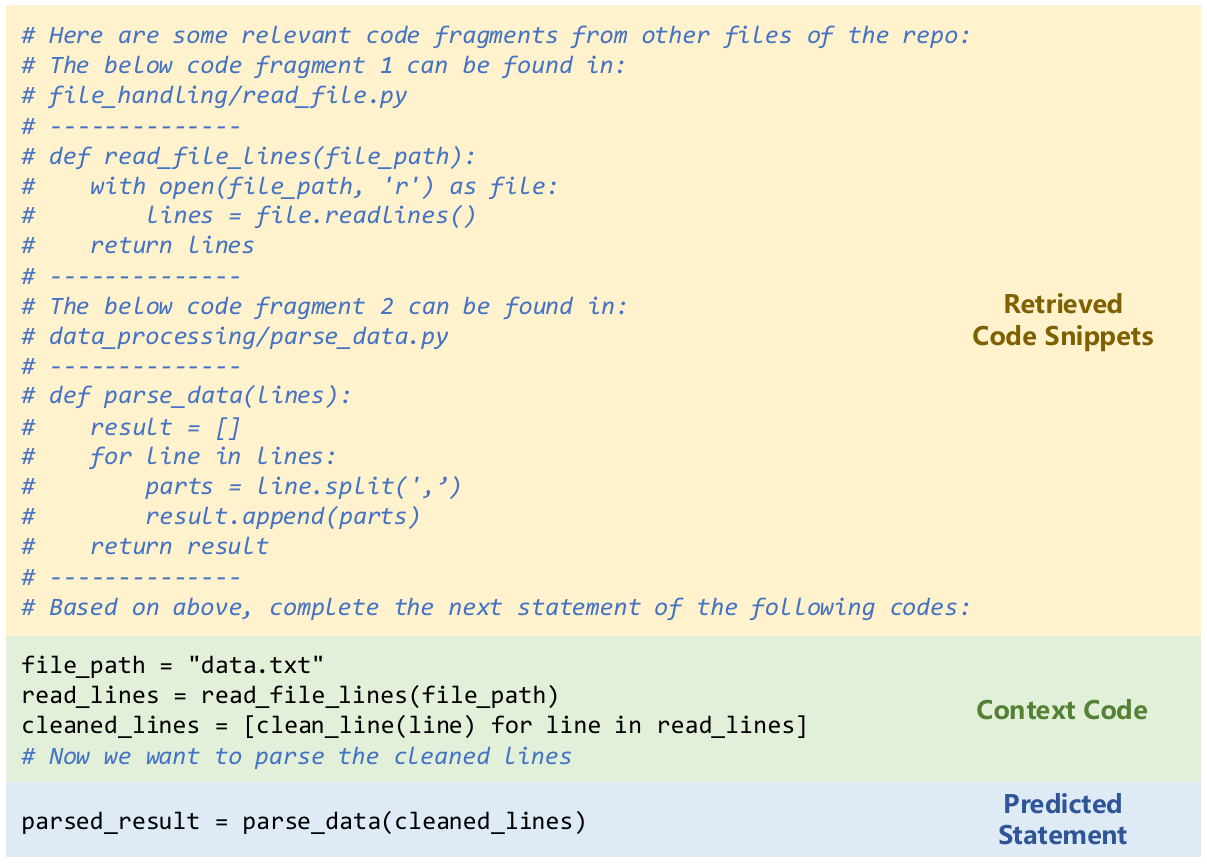}
    \caption{Prompt template used in SelfRACG.}
    \label{fig:prompt_template}
\end{figure*}

Following prior studies~\cite{zhang2023repocoder, wang2024coderag, liu2024graphcoder}, we use the prompt template as presented in Figure~\ref{fig:prompt_template}.

\section{The results of Qwen2.5-Coder-Instruct.}
\label{sec:qwen25}

To further evaluate the generalizability of SelfRACG, we conduct additional experiments using the Qwen2.5-Coder-Instruct~\cite{hui2024qwen2} models, including both 3B and 7B variants. The results are summarized in Table~\ref{tab:qwen2gen}.

Overall, the performance trends observed with Qwen2.5-Coder-Instruct are consistent with those reported in the main experiments using OpenCoder. 
These results further reinforce the robustness and general applicability of SelfRACG across different model backbones.

\begin{table*}[t]
\caption{The code completion experimental results of Qwen2.5-Coder-Instruct.}
\resizebox{\textwidth}{!}{
\begin{tabular}{l|cc|cc|ccc|cc|cc}
\hline
& \multicolumn{2}{c|}{RepoEval-API} & \multicolumn{2}{c|}{RepoEval-Line} & \multicolumn{3}{c|}{RepoEval-Function} & \multicolumn{2}{c|}{CrossCodeEval} & \multicolumn{2}{c}{MEAN} \\
\cline{2-12}
& EM & ES & EM & ES & EM & ES & Pass@1 & EM & ES & EM & ES \\
\hline
\multicolumn{12}{c}{Qwen2.5-Coder-3B-Instruct} \\
\hline
w/o Retrieval & 0.271 & 0.576 & 0.348 & 0.604 & 0.056 & 0.404 & 23.5\% & 0.040 & 0.426 & 0.179 & 0.502 \\
\hline
VanillaRACG &  &  &  &  &  &  & &  & &  &  \\ 
\multicolumn{1}{r|}{\footnotesize\textit{w/ BM25}} & 0.311 & 0.609 & 0.406 & 0.643 & 0.065 & 0.405 & 25.7\% & 0.068 & 0.458 & 0.212 & 0.529 \\
\multicolumn{1}{r|}{\footnotesize\textit{w/ OpenAI-small}} & 0.313 & 0.609 & 0.418 & 0.660 & 0.084 & 0.425 & 26.9\% & 0.086 & 0.465 & 0.225 & 0.540 \\
\multicolumn{1}{r|}{\footnotesize\textit{w/ OpenAI-large}} & 0.317 & 0.611 & 0.426 & 0.667 & 0.084 & 0.432 & 27.2\% & 0.082 & 0.462 & 0.227 & 0.543 \\
\multicolumn{1}{r|}{\footnotesize\textit{w/ NV-Embed-v2}} & \underline{0.343} & 0.623 & 0.436 & 0.669 & 0.084 & 0.425 & 27.9\% & 0.098 & 0.474 & 0.240 & 0.548 \\
\multicolumn{1}{r|}{\footnotesize\textit{w/ GritLM-7B}} & 0.335 & \underline{0.626} & \underline{0.450} & \underline{0.676} & \underline{0.096} & \underline{0.436} & \underline{30.0\%} & \underline{0.103} & \underline{0.480} & \underline{0.246} & \underline{0.555} \\
\hline
SelfRACG & \textbf{0.364} & \textbf{0.647} & \textbf{0.459} & \textbf{0.688} & \textbf{0.115} & \textbf{0.456} & \textbf{32.2\%} & \textbf{0.105} & \textbf{0.484} & \textbf{0.261} & \textbf{0.569} \\
\hline
\multicolumn{12}{c}{Qwen2.5-Coder-7B-Instruct} \\
\hline
w/o Retrieval & 0.274 & 0.575 & 0.343 & 0.602 & 0.056 & 0.404 & 24.8\% & 0.054 & 0.428 & 0.182 & 0.502 \\
\hline 
VanillaRACG &  &  &  &  &  &  & &  & &  &  \\ 
\multicolumn{1}{r|}{\footnotesize\textit{w/ BM25}} & 0.316 & 0.603 & 0.410 & 0.645 & 0.059 & 0.412 & 27.9\% & 0.080 & 0.460 & 0.216 & 0.530 \\
\multicolumn{1}{r|}{\footnotesize\textit{w/ OpenAI-small}} & 0.324 & 0.611 & 0.412 & 0.652 & 0.087 & 0.436 & 28.2\% & 0.098 & 0.474 & 0.230 & 0.543 \\
\multicolumn{1}{r|}{\footnotesize\textit{w/ OpenAI-large}} & 0.321 & 0.605 & 0.427 & 0.663 & 0.090 & \textbf{0.441} & 29.1\% & 0.099 & 0.474 & 0.234 & 0.546 \\
\multicolumn{1}{r|}{\footnotesize\textit{w/ NV-Embed-v2}} & \underline{0.344} & 0.619 & 0.438 & 0.668 & 0.084 & 0.428 & 26.9\% & 0.112 & 0.484 & 0.244 & 0.550 \\
\multicolumn{1}{r|}{\footnotesize\textit{w/ GritLM-7B}} & 0.339 & \underline{0.623} & \underline{0.443} & \underline{0.668} & \underline{0.096} & \underline{0.437} & \underline{30.7\%} & \underline{0.115} & \underline{0.496} & \underline{0.248} & \underline{0.556} \\
\hline
SelfRACG & \textbf{0.371} & \textbf{0.645} & \textbf{0.443} & \textbf{0.668} & \textbf{0.099} & 0.436 & \textbf{31.6\%} & \textbf{0.132} & \textbf{0.499} & \textbf{0.261} & \textbf{0.562} \\
\hline
\end{tabular}
}
\label{tab:qwen2gen}
\end{table*}

\end{document}